# Dynamics influencing ordering of a *s*=3 Ising antiferromagnetic on a triangular lattice

C. Bentham[1], S. George[1,2], J. Poulton[1,3] and GA Gehring[1*]

[1] Department of Physics and Astronomy, University of Sheffield, Sheffield S3 7RH, UK

[2] CERN, Route de Meyrin 385, 1217 Meyrin, Switzerland

[3] Royal School of Mines, Imperial College London

**Abstract**

Antiferromagnetically coupled Ising *s* =3 spins on a triangular lattice are very close to ordering at zero temperature. The low temperature behaviour of a triangular lattice with Ising spins *s* =3 has been simulated by using both Glauber and Kawasaki dynamics. The formation of misfit clusters, which are essential for destabilizing the ordered state, are inhibited by the use of Kawasaki dynamics. The sublattice susceptibilities and the sublattice order parameter are found to depend qualitatively on the dynamics used so that robust ordering occurs when Kawasaki dynamics are employed. The thermal behaviour that is found for the spin model with Kawasaki dynamics gives insight into the observed ordering of side chains seen in a tetraphilic liquid crystal.

- Corresponding author: g.gehring@shef.ac.uk, Tel (44) 114 222 4299, Fax (44) 114 222 3555

[Type here]

**I Introduction**

There has been much interest in the study of spins on regular lattices where antiferromagnetic interactions produce frustration so that the ground state has finite entropy because of disorder on one or all sublattices [1, 2, 3]. Such lattices may display ground states that are partially ordered but still have finite entropy or there may be no long range order. Ising spins on a triangular lattice show both of these phenomena; in all cases the ground state energy is given by $-|J|/s^2$ per spin, where $J$ is the exchange energy, and any ordering is driven by the entropy difference between the different phases. In the limit of large spins there is partial order that leaves 1/3 of the spins free and hence leads to a ground state entropy of $\frac{1}{3}k_B\log_e(2s+1)$ [4]. The exact ground state is known for $s = ½$ and its entropy is considerably larger than $\frac{1}{3}k_B\log_e 2$ [5]. The nature of the ground state can be found as a function of a general spin $s$ from a study of the critical exponent, $\eta$, which describes the decay of the long range spin correlation functions $<s(0)s(r)> \sim r^{-\eta}$ [6]. The phase with long range order becomes stable for $s \sim 3$ and there is another phase that occurs below 3 and is destroyed for $2 < s < 5/2$ that has a Kosterlitz-Thouless transition and a fully disordered phase for $s < 2$ [7]. Recent work [8] has shown that the entropy per spin increases smoothly from the exact value at $s = ½$ of 0.32306 to 0.62331 for $s = 5/2$ which is close to the value for $1/3\log_e 7$ which is 0.64864. There has been much discussion of whether the border of the phase with long range order is exactly at $s = 3$ [6, 7] or whether if the calculations are done very accurately one finds that the border lies just below $s = 3$ so that $s = 3$ itself is actually in the phase with long range order [7]. Hence there is interest in further exploring the model with $s = 3$.

There is another motivation for our work. The ordering of side chains in a liquid crystal can be mapped (with approximations) on to the ordering of a spin $s = 3$ on a triangular lattice with antiferromagnetic interactions [9, 10]. This physical system certainly shows long range order however this system is constrained to follow Kawasaki dynamics. Hence we were motivated to consider the difference between the ordering behaviour of the $s=3$ Ising spin system antiferromagnetically coupled on a triangular lattice using both Kawasaki and Glauber dynamics. In most systems the type of dynamics that is used to drive the system to equilibrium affects only the dynamic critical exponents [10] however because the ordering is driven by the entropy in this case we may hope to see a significant dependence on the type of thermalisation that can occur.

We use Monte Carlo methods with the Metropolis algorithm to model the spin system. In section II a qualitative understanding is given for the way in which the state with long range order may become destabilized. The relevance of the dynamic process, Glauber or Kawasaki, in generating the instability of the ordered phase is described in section III and the simulation results are presented in section IV. Finally the relevance of this work to the ordering of the liquid crystal is discussed in section V and conclusions given in section VI.

**II Qualitative Description of the ordering**

For large spin, $s>3$ the ground state has long range order on two sublattices and the spin on the third sublattice is random taking all values $-s \leq m \leq s$ with equal probability as shown in Figs 1a and 1b where the sublattices are labeled $A, B, C$. For $s<3$ the ground state disorders because of the formation of misfit clusters as shown in Fig 1c [4, 6]. This occurs when the three spins on the random sublattice (C) surrounding a site with spin $m_A=+s$ on the ordered sublattice (A) all take the value $m_C=+s$ so that the originally ordered site on the A sublattice becomes disordered. The energy of the defect cluster is actually no higher than that of the ordered state however the entropy is different. There are two competing contributions to the entropy; there is a configuration entropy because the defect cluster

[Type here]

may be placed in any position on the lattice but also the entropy associated with the sites where the spins originally took random values, $k_B \log_e(2S+1)$, is lost because these sites must now be occupied by a spin $m_C=s$ [4]. This competition means that the ground state is stable for high spin but for low spin, $S$, these defects proliferate and destroy the long range order [4, 6].

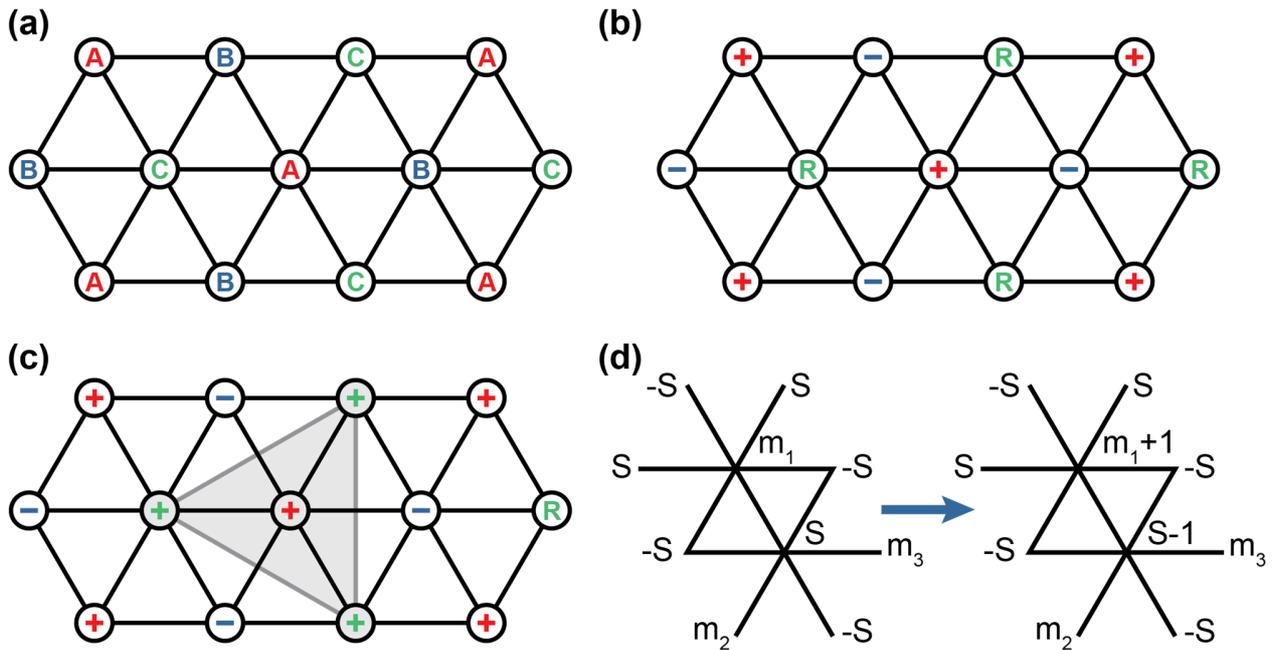

Figure 1. (a) The sublattices for the ordered phase (b). The ordered ground state: the sites labeled $+/-$ are in the states $m=+s/-s$ and the site labeled $R$ is in a random mixed state. (c) A defect cluster: three sites on the random sublattice (C) have taken the value $+s$ leading to an $A$ site, that was originally in the state $+s$ now having an equal number of neighbors that are in $+s$ and $-s$ states so it now becomes random. (d) The bonds whose energies will change when two spins flips occur so $m_1, s$ becomes $m_1+1, s-1$.

**III Glauber and Kawasaki dynamics**

Most Monte Carlo calculations are performed using Glauber dynamics. In this case each spin is allowed to make a flip to another state $m \to m'$ according to the Monte Carlo rules for the Metropolis

[Type here]

algorithm [11]. For the triangular magnet there is thus a probability of $(2s+1)^{-3}$ that three spins on the random sublattice all take the value $+s$.

Kawasaki spin dynamics means that the Monte Carlo procedure is done for a pair of neighboring spins that may be flipped together so that the total z component of spin is conserved, $m, m' \to m+1, m'-1$. In Fig 1d we show a part of the ground state where one of the sites which has taken a random value, $m_1$, has changed to $m_1 +1$ and the neighbouring spin has reduced its z component to $s-1$. The bonds shown are those whose energy will be changed after these spin flips.

The initial, $E_1$, and final, $E_2$, energies and the change in energy are given by:

$$E_1 = -3Js^2 + Js(m_2 + m_3) \qquad E_2 = -J(1+m_1+m_2+m_3) \qquad E_2 - E_1 = -J(1+m_1+m_2+m_3-3S) \; [1]$$

The values of $m_1 +1$, $m_2$ and $m_3$ must each be less than or equal to $s$ hence this process never lowers the energy of this cluster. The energy change is zero only if $m_1 = s-1$ and $m_2 = m_3 = s$ which is exactly the condition for a disordered cluster to form. This shows that in Kawasaki dynamics the disordering of the ground state is strongly inhibited because many of the possible transitions lead to a state with a higher energy and so will be inhibited strongly at low temperatures.

**IV The Monte Carlo Calculations**

We used the Metroplis algorithm for a lattice of 120×120 and used 6000 flips/spin to initialise our system at high temperatures and a further 6300 flips/spin per temperature point recording the lattice configuration over the final 300 flips/spin to avoid hysteresis. The plots shown below in Figure 2 were averaged over 1000 runs for both Glauber and Kawasaki dynamics, in all cases the quantities are plotted against $k_B T / J$.

[Type here]

Calculations were made of the sublattice orderings $\frac{<m_A>}{3}, \frac{<m_B>}{3}, \frac{<m_C>}{3}$ as shown in Figures 2a, b for Kawasaki and Glauber dynamics respectively. These calculations appear to show that in our finite lattice the sublattices were interchanging over a narrow temperature range $4 < k_B T / J < 7$ for Kawasaki dynamics. The interchange appears to go to much lower temperatures for Glauber dynamics and it is possible that the simulation was coming out of thermal equilibrium due to the long relaxation times. Sublattice susceptibilities are a more sensitive probe and these are plotted in figure 2c and these show that all three sublattices have a large (possibly divergent in the thermodynamic limit) susceptibility when Kawasaki dynamics is used. This comes just at the point where the sublattices are beginning to order as seen in figure 2a. There is no sign of an increased susceptibility at this temperature when Glauber dynamics are used as shown in figure 2d and the only divergence appears in the limit of $T \to 0$. This indicates that at $k_B T \sim 5J$ where the fluctuations start appearing in the order parameters as shown in figure 2b there is only short range order which is showing up in our data because of the finite size of the lattice.

Finally the specific heat is shown in plots 2e,f. For both types of dynamics a peak shows up around $k_B T \sim 6J$ as indicated earlier the ground state energy is always lowered by an amount $-|J|s^2$ so at least a Schottky peak is expected. We deduce from the coincidence of the peaks in the sublattice susceptibility and the specific heat that coincide with the onset of finite order parameters that long range order can occur when Kawasaki dynamics are used. The results for the Glauber dynamics are less clear and our results are compatible the model only exhibiting short range order below $k_B T \sim 6J$.

[Type here]

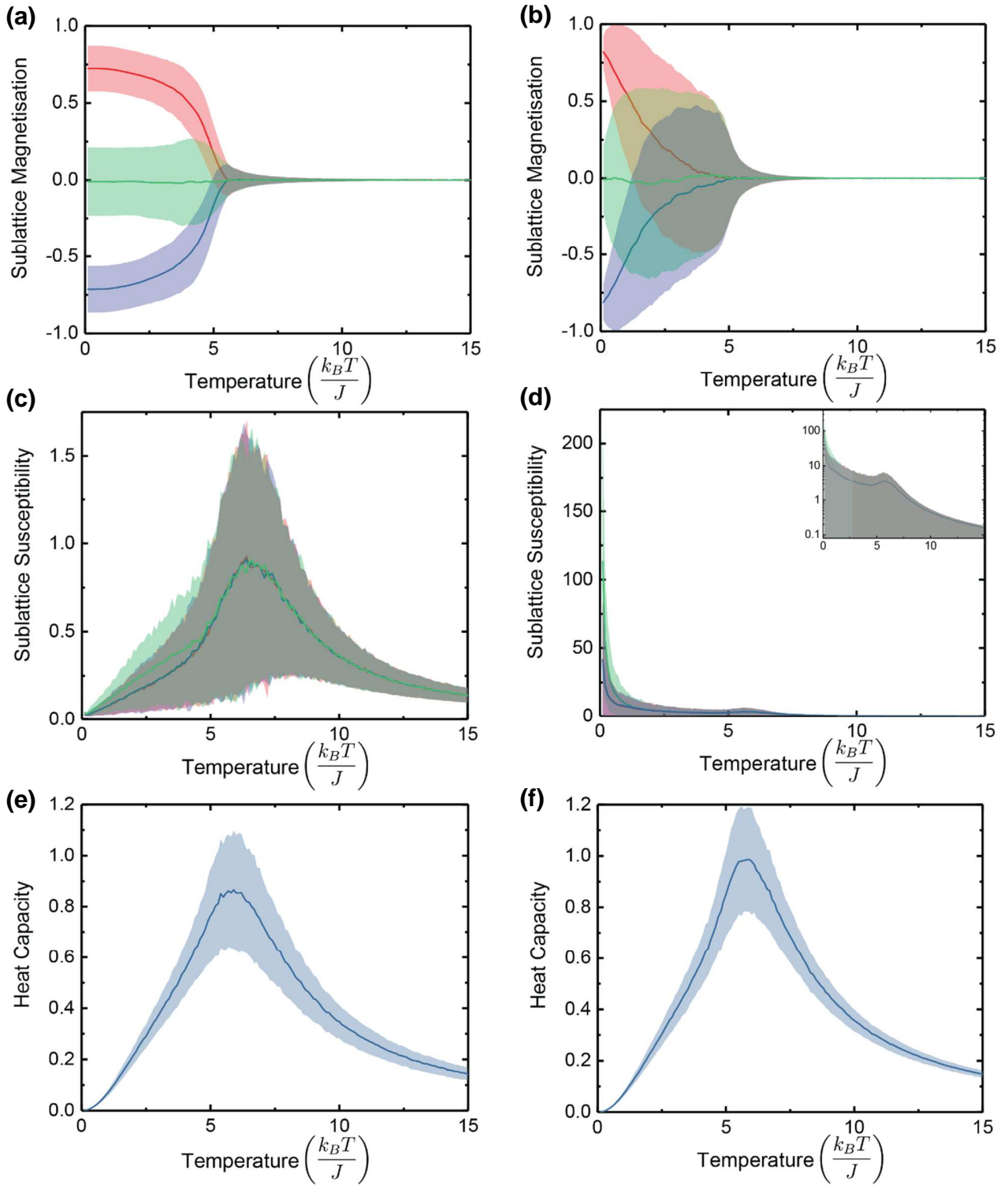

Figure 2. The results of the Monte Carlo simulations plotted against $k_B T/J$ for the sublattice magnetisations (a)/(b) the sublattice susceptibilities (c)/(d) and specific heats (e)/(f) for Kawasaki and Glauber dynamics respectively. A, B, and C sublattices are shown in red, blue, and green, respectively. Shaded areas correspond to standard deviations.

[Type here]

**V Relationship between ordering of side chains and spins on a triangular lattice**

The model of an Ising $s=3$ spin on a triangular lattice with Kawaski dynamics is a good representation of ordering of side chains in a hexagonal liquid crystal lattice [4]. In this case the rigid liquid crystal molecules each with two side chains, one with silicon groups and one with fluorine groups form a honeycomb lattice so that each liquid crystal molecule donates one side chain to the centre of two adjacent hexagons. The Si and F chains repel leading to the fully ordered state is shown in Fig 3a where the Si chains are coloured red and the F chains blue. This differs from the ordered state of the $s=3$ triangle in a very important respect namely that that the mixed sublattice contains The Si and F chains repel but as the hexagons lie on a triangular lattice there is frustration and the fully ordered state mirrors the fully ordered state. The order parameter for the chain problem is given by the difference in the number of Si and F chains, $n_{Si}$, $n_F$, in each hexagon and maps on to the $s=3$ Ising model with $m = \dfrac{n_{Si} - n_F}{2}$.

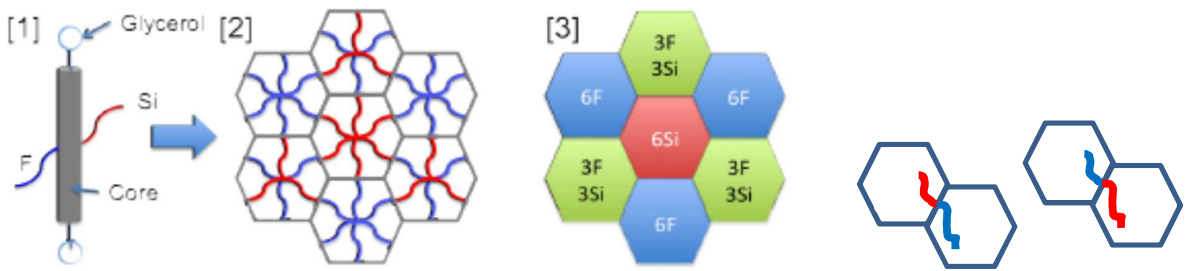

Figure 3: The side chains shown in (a) self-organize into a honeycomb cell (b) with as many side chains of the same type as possible. (c) The system disorders through 180 degree rotations of the side chain pairs.

[Type here]

The effect of disordering involves interchanging the chains on a single liquid crystal molecule so a Si chain interchanged with a F chain in one hexagon and an F chain for a Si chain in the neighboring hexagon. This is an exact rule because the number of Si and F chains is fixed by the initial chemistry. Hence some of the features of the order-disorder transition observed in the liquid crystal may be understood in terms of the $s=3$ spin model on a triangular lattice provided that Kawasaki dynamics are used in the modeling.

**VI Conclusions**

The model of spins $s=3$ on a triangular lattice is very close to ordering at low temperatures and the entropy plays a crucial role as in other 'order-by-disorder' transitions [12]. We have shown that the use of Kawasaki dynamics inhibits the formation of misfit clusters and hence favours the formation of the ordered state. This is an unusual scenario as it is well known that the use of dynamics with a conserved magnetisation usually influences the dynamic critical exponents but not the thermodynamics [13]. The simulations in this case do show a striking dependence on whether Kawaski or Glauber dynamics was used. This occurs because the transition is driven by the entropy in this case. The ordering shown in the case when Kawasaki dynamics are used gives insight as to why the side chains in the liquid crystal show an order-disorder transition.

**Acknowledgements**

We should like to thank Dr. Zeng and Professor Ungar for many interesting discussions and Professor Henley for very helpful correspondence.

**References**

1. Liebmann, R. Statistical Mechanics of Periodic Frustrated Ising Systems Springer, Berlin (1986)
2. Collins, M.F. and Petrenko, O.A.: Can. J Phys **75**, 605 (1997)

[Type here]

3. Diep, H.T. (ed.) Frustrated spin Systems Imperial college Press (2013)

4. Nagai, Ojiro, Miyashita, Seiji and Horiguchi, Tsuyoshi: Phys Rev 47, 202 (1993)

5. Wannier, G.H., Phys. Rev. **79**, 357 (1950); Phys. Rev. B **7**, 5017 (1973)

6. Lipowski, A., Horiguchi, T. and Lipowska, D.: Phys Rev Lett. **74**, 3888 (1995)

7. Zeng, C and Henley C.L.: Phys Rev B **55**, 14935 (1997)

8. X. Zeng, X.. Kieffer, R.,. Glettner, B.,. Numberger, C., Liu, F., Pelz, K., Prehm, M., Baumeister, U., Hahn, H., Lang, H., Gehring, G. A., Weber, C. H. M. , Hobbs, J. K. Tschierske, C., Ungar G.: Science, **331**, 1302 (2011)

9. George, S., Bentham, C., Zeng, X., Ungar G. and Gehring G.A.: Phys Rev E **95**, 062126 (2017)

10. Žukovič Milan: Eur. Phys. J. B **86**, 238 (2013);

11. Hohenberg P.C.and Halperin B. I. : Rev. Mod. Phys. 49, 435 (1977).

12. Metropolis, N., Rosenbluth A.W. and. Rosenbluth M.N: J. Chem. Phys. **21**, 1087 (1953)

13. Moessner R. Can. J. Phys. **79**, 1283 (2001)